\documentclass[10pt,a4paper,twocolumn]{article}
\usepackage{graphicx}

\begin{document}
\title{The novel aspect of hydrogen atom: the fine structure and spin can be derived by
single component wavefunction}
\author{\small H. Y. Cui\\
\small Department of Physics, Beihang University\\
\small Beijing, 100083, China, E-mail: hycui@buaa.edu.cn}
\date{\small \today}

\maketitle

\begin{abstract}
\small
The fine structure of hydrogen energy was calculated by using 
the usual momentum-wavefunction relation directly, rather than
establishing the well-known Dirac wave equation. As the results,
the energy levels are completely the same as that of the Dirac
wave equation, while the wavefunction is single component that is
quite different from Dirac's four component wavefunction, the
most important thing is that the present calculation brings out
electronic spin in a new way which has never been reported and
indicates that the electronic spin is a kind of orbital motion.
\end{abstract}

\section{Introduction}

As we know, the fine structure and electronic spin in hydrogen atom can be
exactly calculated by using the Dirac wave equation which contains four
component wavefunction, the work is a milestone in physics. Up to now, it
has been believed that electronic spin must be theocretically specified by
Dirac's four component wavefunction, spin concept has been developed as
one of the most important properties of elementary particles in high energy
physics.

Consider a particle of rest mass $m$ and charge $q$ moving in an inertial
frame of reference with relativistic 4-vector velocity $u_{\mu }$, it
satisfies\cite{Harris}

\begin{equation}
u_{\mu }u_{\mu }=-c^{2}  \label{s1}
\end{equation}%
where there is not distinction between covariant and contravariant
components in the Cartesian coordinate system. Eq.(\ref{s1}) is just the
relativistic energy-momentum relation when multiplying it by squared mass.
Let $A_{\mu }$ denote vector potential of electromagnetic field, 
substituting the momentum-wavefunction relation\cite{Harris1}

\begin{equation}
mu_{\mu }=\frac{1}{\psi }(-i\hbar \partial _{\mu }-qA_{\mu })\psi
\label{s2}
\end{equation}%
into Eq.(\ref{s1}), we obtain a new quantum wave equation with single
component wavefunction

\begin{equation}
\lbrack (-i\hbar \partial _{\mu }-qA_{\mu })\psi ][(-i\hbar \partial _{\mu
}-qA_{\mu })\psi ]=-m^{2}c^{2}\psi ^{2}  \label{s3}
\end{equation}%
Note that taking the right side of Eq.(\ref{s2}) as momentum-operator to
replace momentum that appears in a physical equation is a traditional
usage, but in this paper we directly use the right side of Eq.(\ref{s2}) as
momentum as we have done in Eq.(\ref{s3}). Please note that Eq.(\ref{s3}%
) is not the Klein-Gordon wave equation.

In the recent years, H. Y. Cui\cite{Cui,Cui1} has reported some significant
results of Eq.(\ref{s3}) for certain physical systems. In this paper I point
out that the fine structure of hydrogen energy can also be calculated by
using Eq.(\ref{s3}), while its wavefunction has single component in contrast
with Dirac's wavefunction, the spin effect of electron is also revealed by
Eq.(\ref{s3}) when the hydrogen atom is in a magnetic field, the present
calculation totally does not need multi-component wavefunction for
specifying electronic spin, and indicates that we should re-recognize spin.

\section{The fine structure of hydrogen energy}


In the following, we use Gaussian units, and use $m_e$ to denote the rest
mass of electron.

In a spherical polar coordinate system $(r,\theta ,\varphi ,ict)$, the
nucleus of hydrogen atom provides a spherically symmetric potential $V(r)=e/r
$ for the electron motion. The wave equation (\ref{s3}) for the hydrogen
atom in the energy eigenstate $\psi (r,\theta ,\varphi )e^{-iEt/\hbar }$ may
be written in the spherical coordinates:

\begin{equation}
\frac{m_{e}^{2}c^{2}}{\hbar ^{2}}\psi ^{2}=(\frac{\partial \psi }{\partial r}%
)^{2}+(\frac{1}{r}\frac{\partial \psi }{\partial \theta })^{2}+(\frac{1}{%
r\sin \theta }\frac{\partial \psi }{\partial \varphi })^{2}+\frac{1}{\hbar
^{2}c^{2}}(E+\frac{e^{2}}{r})^{2}\psi ^{2} \nonumber \\
 \label{s4}
\end{equation}%
By substituting $\psi =R(r)X(\theta )\phi (\varphi )$, we separate the above
equation into

\begin{eqnarray}
(\frac{\partial \phi }{\partial \varphi })^{2}+\kappa \phi ^{2} &=&0\nonumber \\
\label{s5} \\
(\frac{\partial X}{\partial \theta })^{2}+[\lambda -\frac{\kappa }{\sin
^{2}\theta }]X^{2} &=&0  \nonumber \\
\label{s6} \\
(\frac{\partial R}{\partial r})^{2}+[\frac{1}{\hbar ^{2}c^{2}}(E+\frac{e^{2}%
}{r})^{2}-\frac{m_{e}^{2}c^{2}}{\hbar ^{2}}-\frac{\lambda }{r^{2}}]R^{2} &=&0\nonumber \\
\label{s7}
\end{eqnarray}%
where $\kappa $ and $\lambda $ are constants introduced for the separation.
Eq.(\ref{s5}) can be solved immediately, with the requirement that $\phi
(\varphi )$ must be a periodic function, we find its solution given by

\begin{equation}
\phi =C_{1}e^{\pm i\sqrt{\kappa }\varphi }=C_{1}e^{im\varphi }\qquad m=\pm
\sqrt{\kappa }=0,\pm 1,...  \label{s8}
\end{equation}%
where $C_{1}$ is an integral constant.

It is easy to find the solution of Eq.(\ref{s6}), it is given by

\begin{equation}
X(\theta )=C_{2}e^{\pm i\int \sqrt{\lambda -\frac{m^{2}}{\sin ^{2}\theta }}%
d\theta }  \label{s9}
\end{equation}%
where $C_{2}$ is an integral constant. The requirement of periodic function
for $X$ demands

\begin{equation}
\int\nolimits_{0}^{2\pi }\sqrt{\lambda -\frac{m^{2}}{\sin ^{2}\theta }}%
d\theta =2\pi k\qquad k=0,1,2,...  \label{s10}
\end{equation}%
This complex integration is evaluated (see the appendix for the details), we
get

\begin{equation}
\int\nolimits_{0}^{2\pi }\sqrt{\lambda -\frac{m^{2}}{\sin ^{2}\theta }}%
d\theta =2\pi (\sqrt{\lambda }-|m|)  \label{s11}
\end{equation}%
thus, we obtain

\begin{equation}
\sqrt{\lambda }=k+|m|  \label{s12}
\end{equation}%
We rename the integer $\lambda $ as $j^{2}$ for a convenience in the
following, i.e. $\lambda =j^{2}$.

The solution of Eq.(\ref{s7}) is given by

\begin{equation}
R(r)=C_{3}e^{\pm \frac{i}{\hbar c}\int \sqrt{(E+\frac{e^{2}}{r}%
)^{2}-m_{e}^{2}c^{4}-\frac{\lambda \hbar ^{2}c^{2}}{r^{2}}}dr}  \label{s13}
\end{equation}%
where $C_{3}$ is an integral constant. The requirement that the radical
wavefunction forms a \textquotedblright standing wave\textquotedblright\ in
the range from $r=0$ to $r=\infty $ demands

\begin{eqnarray}
\frac{1}{\hbar c}\int\nolimits_{0}^{\infty }\sqrt{(E+\frac{e^{2}}{r}%
)^{2}-m_{e}^{2}c^{4}-\frac{\lambda \hbar ^{2}c^{2}}{r^{2}}}dr=\pi s \nonumber \\
s=0,1,2,...  \label{s14}
\end{eqnarray}%
This complex integration is evaluated (see the appendix for the details), we
get

\begin{eqnarray}
\frac{1}{\hbar c}\int\nolimits_{0}^{\infty }\sqrt{(E+\frac{e^{2}}{r}%
)^{2}-m_{e}^{2}c^{4}-\frac{j^{2}\hbar ^{2}c^{2}}{r^{2}}}dr\nonumber \\
=\frac{\pi E\alpha}{\sqrt{m_{e}^{2}c^{4}-E^{2}}}-\pi \sqrt{j^{2}-\alpha ^{2}}  \label{s15}
\end{eqnarray}%
where $\alpha =e^{2}/\hbar c$ is known as the fine structure constant.


Form the last Eq.(\ref{s14}) and Eq.(\ref{s15}), we obtain the energy levels
given by

\begin{equation}
E=m_{e}c^{2}\left[ 1+\frac{\alpha ^{2}}{(\sqrt{j^{2}-\alpha ^{2}}+s)^{2}}%
\right] ^{-\frac{1}{2}}  \label{s16}
\end{equation}%
where $j=\sqrt{\lambda }=k+|m|$. Because $j\neq 0$ in Eq.(\ref{s16}), we
find $j=1,2,3...$.

The result, Eq.(\ref{s16}), is completely the same as that in the
calculation of the Dirac wave equation\cite{Schiff} for the hydrogen atom,
it is just the \textbf{fine structure of hydrogen energy}.

\section{The electronic spin}


If we put the hydrogen atom into an external uniform magnetic field $B$
which is along the $z$ axis with the vector potential $(A_{r},A_{\theta
},A_{\varphi })=(0,0,\frac{1}{2}r\sin \theta B)$, then according to Eq.(\ref%
{s3}) the wave equation is given by

\begin{eqnarray}
\frac{m_e^2c^2}{\hbar ^2}\psi ^2 &=&(\frac{\partial \psi }{\partial r}%
)^2+(\frac 1r\frac{\partial \psi }{\partial \theta })^2+\frac 1{\hbar
^2c^2}(E+\frac{e^2}r)^2\psi ^2  \nonumber \\
&&+(\frac 1{r\sin \theta }\frac{\partial \psi }{\partial \varphi }-\frac{%
er\sin \theta B}{i2c\hbar }\psi )^2  \label{d1}
\end{eqnarray}
By substituting $\psi =R(r)X(\theta )\phi (\varphi )$, we separate the above
equation into

\begin{eqnarray}
\frac{\partial \phi }{\partial \varphi }+\kappa \phi &=&0  \nonumber \\
\label{d2} \\
(\frac{\partial X}{\partial \theta })^{2}+[(\frac{-\kappa }{\sin \theta }-%
\frac{e\sin \theta r^{2}B}{i2c\hbar })^{2}+\xi (r)]X^{2} &=&0  \nonumber \\
\label{d3} \\
(\frac{\partial R}{\partial r})^{2}+[\frac{1}{\hbar ^{2}c^{2}}(E+\frac{e^{2}%
}{r})^{2}-\frac{m_{e}^{2}c^{2}}{\hbar ^{2}}-\frac{\xi (r)}{r^{2}}]R^{2} &=&0\nonumber \\
\label{d4}
\end{eqnarray}%
where we have used the unknown constant $\kappa $ and function $\xi (r)$ to
connect these separated equations. Eq.(\ref{d2}) has the solution

\begin{equation}
\phi =C_{1}e^{im\varphi },\qquad \kappa =im,\qquad m=0,\pm 1,\pm 2,...
\label{d5}
\end{equation}%
Expanding Eq.(\ref{d3}) and neglecting the term $O(B^{2})$, we find a
constant term $-\frac{mer^{2}B}{c\hbar }$ in it, by moving this term into
Eq.(\ref{d4}) through $\xi (r)=\lambda +\frac{mer^{2}B}{c\hbar }$, we obtain

\begin{eqnarray}
(\frac{\partial X}{\partial \theta })^{2}+[\lambda -\frac{m^{2}}{\sin
^{2}\theta }]X^{2} &=&0 \nonumber \\
 \label{d6} \\
(\frac{\partial R}{\partial r})^{2}+[\frac{1}{\hbar ^{2}c^{2}}(E+\frac{e^{2}%
}{r})^{2}-\frac{m_{e}^{2}c^{2}}{\hbar ^{2}}-\frac{meB}{c\hbar }-\frac{%
\lambda }{r^{2}}]R^{2} &=&0 \nonumber \\
 \label{d7}
\end{eqnarray}%
The above two equations are the same as Eq.(\ref{s6}) and Eq.(\ref{s7}),
except for the additional constant term $-meB/c\hbar $. After the similar
calculation as the preceding section, we obtain the energy levels of
hydrogen atom in the magnetic field given by

\begin{equation}
E=\sqrt{m_{e}^{2}c^{4}+mec\hbar B}\left[ 1+\frac{\alpha ^{2}}{(\sqrt{%
j^{2}-\alpha ^{2}}+s)^{2}}\right] ^{-\frac{1}{2}}  \label{d8}
\end{equation}


In the usual spectroscopic notation of quantum mechanics, four quantum
numbers: $n$, $l$, $m_l$ and $m_s$ are used to specify the state of an
electron in an atom. After the comparison, we get the relations between the
usual notation and our notation.

\begin{eqnarray}
n &=&j+s,\quad s=0,1,...;j=1,2,....  \label{d10} \\
l &=&j-1,  \label{d11} \\
\quad \max (m_{l}) &=&\max (m)-1  \label{d12}
\end{eqnarray}%
We find that $j$ takes over $1,2,...,n$; for a fixed $j$ (or $l$), $m$ takes
over $-(l+1),-l,...,0,...,l,l+1$. In the present work, spin quantum number
is absent.


According to Eq.(\ref{d8}), for a fixed $(n,l)$, equivalent to $(n,j=l+1)$,
the energy level of hydrogen atom will split into $2l+3$ energy levels in
the magnetic field, given by

\begin{equation}
E=(m_{e}c^{2}+\frac{me\hbar B}{2m_{e}c})\left[ 1+\frac{\alpha ^{2}}{(\sqrt{%
j^{2}-\alpha ^{2}}+s)^{2}}\right] ^{-\frac{1}{2}}+O(B^{2})  \nonumber \\
\label{d13}
\end{equation}%
Considering $m=-(l+1),-l,...,0,...,l,l+1$, this effect is equivalent to the
usual Zeeman splitting in the usual quantum mechanics given by

\begin{equation}
E=E_{nl}+\frac{(m_{l}\pm 1)e\hbar B}{2m_{e}c}  \label{d14}
\end{equation}%
But our work works on it without \textbf{spin} concept, the
so-called spin effect has been revealed by Eq.(\ref{d13}) without
spin concept, this result indicates that electronic spin is a kind
of orbital motion.

In Stern-Gerlach experiment, the angular momentum of ground state of
hydrogen atom is presumed to be zero according to the usual quantum
mechanics, thus ones need make use of the spin. But in the present
calculation, the so-called spin has been merged with the orbital motion of
the electron.

Recalling that the spin is a mysterious concept even for very
learned persons, we have come to this point: we can understand the
quantum mechanics without the well-established spin concept, can't
we? without multi-component wavefunction, can't we? Bear in mind
that simplicity is always a merit for the physics.

\section{Conclusion}

Using equations

\begin{eqnarray}
(mu_{\mu }+qA_{\mu })\psi &=&-i\hbar \partial _{\mu }\psi  \label{e1} \\
u_{\mu }u_{\mu } &=&-c^{2}  \label{e2}
\end{eqnarray}%
by eliminating $u_{\mu }$, the above equations can be written as

\begin{equation}
-m^{2}c^{2}\psi ^{2}=[(-i\hbar \partial _{\mu }-qA_{\mu })\psi][
(-i\hbar\partial _{\mu }-qA_{\mu })\psi ]  \label{e3}
\end{equation}%
Using this quantum wave equation, the fine structure of hydrogen atom energy
is calculated, as the results, the energy levels are given by

\begin{eqnarray}
E &=&m_{e}c^{2}\left[ 1+\frac{\alpha ^{2}}{(\sqrt{j^{2}-\alpha ^{2}}+n)^{2}}%
\right] ^{-\frac{1}{2}}  \label{e4} \\
j &=&1,2,...;n=0,1,...  \nonumber
\end{eqnarray}%
the result is completely the same as that in the calculation of
the Dirac wave equation for the hydrogen atom, while the
wavefunction is quite different from Dirac's wavefunction.
Besides, the present calculation brings out spin nature in a new
way, indicating that electronic spin is a kind of orbital motion.
The present calculation is characterized by using the usual 
momentum-wavefunction relation directly, it provides an insight into
the foundations of quantum mechanics.

\appendix

\section{Appendix: The evaluations of the integrations}

In this appendix we give out the evaluations of the integrations 
appeared in the preceding section, i.e.Eq.(\ref{s11}) and Eq.(\ref{s15}).

\subsection{wave-attenuating boundary condition}

\begin{figure}[htb]
\centering \includegraphics[bb=120 580 300 700,clip]{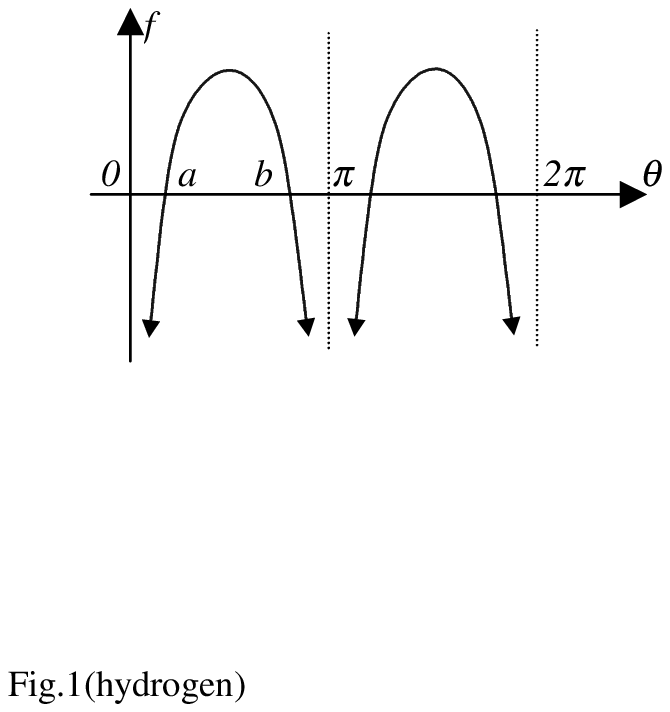}
\caption{The function has sign-changed points $a$ and $b$.}
\label{fig1}
\end{figure}

Consider the integrand in Eq.(\ref{s11}), it is a multiple-valued function,
may be written as

\begin{equation}
\sqrt{\lambda -\frac{m^{2}}{\sin ^{2}\theta }}=\sqrt{f(\theta )}\qquad
f(\theta )=\lambda -\frac{m^{2}}{\sin ^{2}\theta }  \label{a1}
\end{equation}%
Suppose that $\lambda $ is a real positive number, the function $f(\theta )$
may be divided into the three regions: $(0,a)$, $(a,b)$, and $(b,\pi )$, where $%
a $ and $b$ are the turning points at where the function $f(\theta )$
changes its sign, as shown in Fig.\ref{fig1}. we find

\begin{eqnarray}
\int\nolimits_{0}^{a}\sqrt{f(\theta )}d\theta &=&\int\nolimits_{0}^{a}\sqrt{%
-|f(\theta )|}d\theta =\pm i\int\nolimits_{0}^{a}\sqrt{|f(\theta )|}d\theta \nonumber \\
&=&\pm iA \label{a2} \\
\int\nolimits_{a}^{b}\sqrt{f(\theta )}d\theta &=&\int\nolimits_{a}^{b}\sqrt{%
|f(\theta )|}d\theta =B  \nonumber \\
\label{a3} \\
\int\nolimits_{b}^{\pi }\sqrt{f(\theta )}d\theta &=&\int\nolimits_{b}^{\pi }%
\sqrt{-|f(\theta )|}d\theta =\pm iA  \nonumber \\
\label{a4}
\end{eqnarray}%
where $A$ and $B$ are two real numbers, then the integration of Eq.(\ref{s11}%
) has three possible solutions given by

\begin{equation}
\int\nolimits_{0}^{2\pi }\sqrt{\lambda -\frac{m^{2}}{\sin ^{2}\theta }}%
d\theta =\left\{
\begin{array}{l}
2(B+2iA) \\
2B \\
2(B-2iA)%
\end{array}%
\right.  \label{a5}
\end{equation}%
The second branch of this result is reasonable, because only it can fulfil
the requirement that the wavefunction is periodic function of $\theta $. The
multiple-valued result arises from that $\sqrt{-|f(\theta )|}=\pm i\sqrt{%
|f(\theta )|}$, like $\sqrt{-5}=\pm i\sqrt{5}$.

\begin{figure}[htb]
\centering \includegraphics[bb=110 565 305 715,clip]{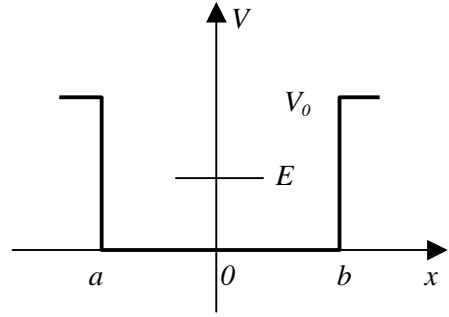}
\caption{A finitely deep potential well that has two turning points $a$ and $%
b$.}
\label{fig2}
\end{figure}

How to determine the sign of the multiple-valued function reasonably? Let us
turn to our experience that we have had in the usual quantum mechanics.
Consider the motion of a particle in a finitely deep potential well as shown
in Fig.\ref{fig2}, there are also two turning points $a$ and $b$. If the
particle moves over the turning point $a$ or $b$ for $E<V_{0}$ (bound
states), its momentum will become imaginary $\pm i|p|$  with  uncertain
sign. As we know in the usual quantum mechanics the wavefunction is given by

\begin{equation}
\psi (x)=\left\{
\begin{array}{ll}
De^{-i\frac{i|p|}{\hbar }x} & x<a \\
G\sin (\frac{|p_{_{0}}|}{\hbar }x+\delta ) & a<x<b \\
De^{-i\frac{-i|p|}{\hbar }x} & x>b%
\end{array}%
\right.  \label{a6}
\end{equation}%
This has involved our experience in which we have taken plus sign for the imaginary
momentum in $x<a$ and minus sign in $x>b$, to satisfy the so-called \textbf{%
wave-attenuating boundary condition} for the regions over the turning points.

In the followings, we use this wave-attenuating boundary condition to determine
the sign of double-valued imaginary momentum: take plus sign in the region
over the left turning point, whereas take minus sign in the region over the
right turning point.

\subsection{integration 1}

To apply the wave-attenuating boundary condition to the following
wavefunction

\begin{equation}
X(\theta )=C_{2}e^{-i\int \sqrt{\lambda -\frac{m^{2}}{\sin ^{2}\theta }}%
d\theta }  \label{a7}
\end{equation}%
due to wave-attenuating for the turning points, the integrand must choose
the signs as

\begin{eqnarray}
\int\nolimits_{0}^{a}\sqrt{\lambda -\frac{m^{2}}{\sin ^{2}\theta }}d\theta
&=&+i\int\nolimits_{0}^{a}\sqrt{|f(\theta )|}d\theta =iA  \label{a8} \\
\int\nolimits_{a}^{b}\sqrt{\lambda -\frac{m^{2}}{\sin ^{2}\theta }}d\theta
&=&\int\nolimits_{a}^{b}\sqrt{f(\theta )}d\theta =B  \label{a9} \\
\int\nolimits_{b}^{\pi }\sqrt{\lambda -\frac{m^{2}}{\sin ^{2}\theta }}%
d\theta &=&-i\int\nolimits_{b}^{\pi }\sqrt{|f(\theta )|}d\theta =-iA
\label{a10}
\end{eqnarray}%
thus the integration may have a real solution, actually it may be written as

\begin{equation}
\int\nolimits_{0}^{2\pi }\sqrt{\lambda -\frac{m^{2}}{\sin ^{2}\theta }}%
d\theta =2\int\nolimits_{a}^{b}\sqrt{\lambda -\frac{m^{2}}{\sin ^{2}\theta }}%
d\theta =2B  \label{a11}
\end{equation}

\begin{figure}[htb]
\centering \includegraphics[bb=160 580 300 705,clip]{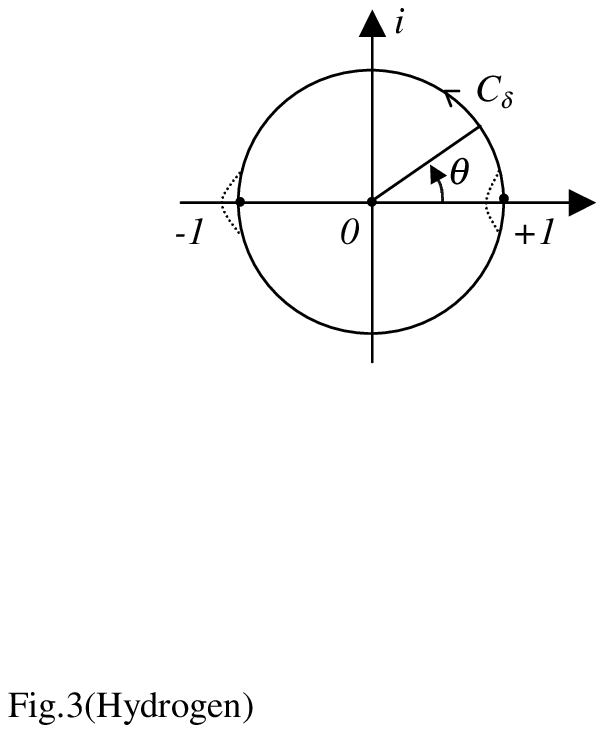}
\caption{Unit circle contour for evaluating integral}
\label{fig3}
\end{figure}

In order to evaluate the definite integral of Eq.(\ref{a11}), we make use of
contour integral in complex plane\cite{Marsden}. Consider a contour $%
C_{\delta }$ which is a unit circle around zero, as shown in Fig.\ref{fig3},
using $z=e^{i\theta }$, we have

\begin{eqnarray}
I_{1} &=&\int\nolimits_{0}^{2\pi }\sqrt{\lambda -\frac{m^{2}}{\sin
^{2}\theta }}d\theta =\int\nolimits_{C_{\delta }}\sqrt{\lambda +\frac{%
4m^{2}z^{2}}{(z^{2}-1)^{2}}}\frac{dz}{iz}  \nonumber \\
&=&\int\nolimits_{C_{\delta }}\frac{\sqrt{\lambda (z^{2}-1)^{2}+4m^{2}z^{2}}%
}{\pm (z^{2}-1)}\frac{dz}{iz}  \label{a12}
\end{eqnarray}%
As we have known that $\sqrt{f(\theta )}|_{\theta =\pi /2,or\theta =3\pi /2}=%
\sqrt{\lambda -m^{2}}$, substituting $z=i$ or $z=-i$ into the above
integrand, we find the integrand must take the minus sign. Thus

\begin{equation}
I_{1}=\int\nolimits_{C_{\delta }}\frac{\sqrt{\lambda
(z^{2}-1)^{2}+4m^{2}z^{2}}}{-(z^{2}-1)}\frac{dz}{iz}  \label{a13}
\end{equation}%
For scrutinizing the sign of the integrand over the turning points, we have

\begin{eqnarray}
\sqrt{f(\theta )} &=&\frac{\sqrt{\lambda (z^{2}-1)^{2}+4m^{2}z^{2}}}{%
-(z^{2}-1)}\nonumber \\
&=&\frac{\sqrt{\lambda (z^{2}-1)^{2}+4m^{2}z^{2}}}{-(2iz)(z^{2}-1)/(2iz)}  \nonumber \\
&=&\frac{\sqrt{\lambda (z^{2}-1)^{2}+4m^{2}z^{2}}}{-(2iz)\sin \theta }\nonumber \\
&=&i\frac{\sqrt{\lambda (z^{2}-1)^{2}+4m^{2}z^{2}}}{2z\sin \theta }  \label{a14}
\end{eqnarray}%
we find that the integrand takes plus sign over the left turning point $%
(\theta \rightarrow 0+,z\rightarrow 1)$ and minus sign over the right
turning point $(\theta \rightarrow \pi -,z\rightarrow -1)$, in accordance
with the sign requirement of Eq.(\ref{a8}) and (\ref{a10}).

Continue our calculation, we have

\begin{eqnarray}
I_{1} &=&\int\nolimits_{C_{\delta }}\frac{\sqrt{\lambda
(z^{2}-1)^{2}+4m^{2}z^{2}}}{-(z^{2}-1)}\frac{dz}{iz}  \nonumber \\
&=&\int\nolimits_{C_{\delta }}(\frac{1}{z}-\frac{1/2}{z-1}-\frac{1/2}{z+1})%
\sqrt{\lambda (z^{2}-1)^{2}+4m^{2}z^{2}}\frac{dz}{i}  \nonumber \\
\label{a16}
\end{eqnarray}%
Now we find that the integrand has the poles at $z=0$ and $z=\pm 1$. We let
the counter $C_{\delta }$ pass by the pole $z=+1$ through the interior of
the unite circle, as indicated by the dash line in Fig.\ref{fig3}, likewise,
let the counter $C_{\delta }$ pass by the pole $z=-1$ through the exterior
of the unite circle. The deformation made for $C_{\delta }$ has no influence
on the integration value because the left deformation cancels the right
deformation in the integration due to the opposite signs of the integrand
near the left and right poles. Let $C_{\delta }^{\prime }$ denote the
deformed counter, we continue the calculation by using residue theorem.

\begin{eqnarray}
I_{1} &=&\int\nolimits_{C_{\delta }^{\prime }}(\frac{1}{z}-\frac{1/2}{z-1}-%
\frac{1/2}{z+1})\sqrt{\lambda (z^{2}-1)^{2}+4m^{2}z^{2}}\frac{dz}{i}
\nonumber \\
&=&\int\nolimits_{C_{\delta }^{\prime }}\frac{1}{z}\sqrt{\lambda
(z^{2}-1)^{2}+4m^{2}z^{2}}\frac{dz}{i}  \nonumber \\
&&-\int\nolimits_{C_{\delta }^{\prime }}\frac{1/2}{z-1}\sqrt{\lambda
(z^{2}-1)^{2}+4m^{2}z^{2}}\frac{dz}{i}  \nonumber \\
&&-\int\nolimits_{C_{\delta }^{\prime }}\frac{1/2}{z+1}\sqrt{\lambda
(z^{2}-1)^{2}+4m^{2}z^{2}}\frac{dz}{i}  \nonumber \\
&=&\int\nolimits_{C_{\delta }^{\prime }}\frac{\sqrt{\lambda }+O(z^{2})}{z}%
\frac{dz}{i}-\int\nolimits_{C_{\delta }^{\prime }}\frac{|m|+O(z^{2}-1)}{z+1}%
\frac{dz}{i}  \nonumber \\
&=&2\pi (\sqrt{\lambda }-|m|)  \label{a17}
\end{eqnarray}

\subsection{Integration 2}

\begin{figure}[htb]
\centering \includegraphics[bb=155 585 320 715,clip]{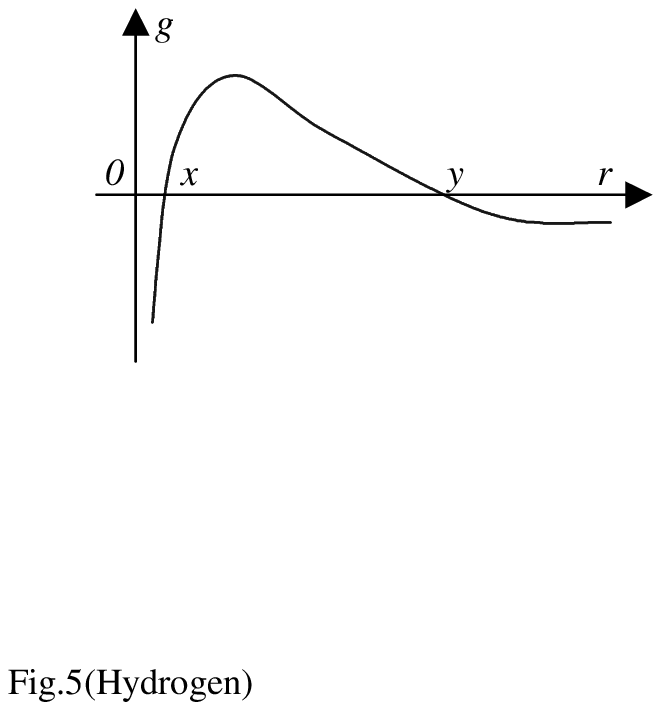}
\caption{Contour for evaluating integral.}
\label{fig4}
\end{figure}

To apply the wave-attenuating boundary condition to the following
wavefunction

\begin{equation}
R(r)=C_{3}e^{-\frac{i}{\hbar c}\int \sqrt{(E+\frac{e^{2}}{r}%
)^{2}-m_{e}^{2}c^{4}-\frac{\lambda \hbar ^{2}c^{2}}{r^{2}}}dr}  \label{a18}
\end{equation}%
where it has also two turning points $x$ and $y$ from $r=0$ to $r=\infty $
when $E^{2}<m_{e}^{2}c^{4}$ (bound states), as shown in Fig.\ref{fig4} where

\begin{equation}
g(r)=(E+\frac{e^{2}}{r})^{2}-m_{e}^{2}c^{4}-\frac{j^{2}\hbar ^{2}c^{2}}{r^{2}%
}  \label{a19}
\end{equation}%
we take the following signs for its asymptotic behavior, i.e.
\begin{eqnarray}
\sqrt{g(r)}|_{r\rightarrow 0} &=&\sqrt{e^{4}-j^{2}\hbar ^{2}c^{2}}/r=i\sqrt{%
j^{2}\hbar ^{2}c^{2}-e^{4}}/r  \nonumber \\
\label{a20} \\
\sqrt{g(r)}|_{r\rightarrow \infty } &=&\sqrt{E^{2}-m_{e}^{2}c^{4}}=-i\sqrt{%
m_{e}^{2}c^{4}-E^{2}}  \nonumber \\
\label{a21}
\end{eqnarray}

\begin{figure}[htb]
\centering \includegraphics[bb=130 565 330 705,clip]{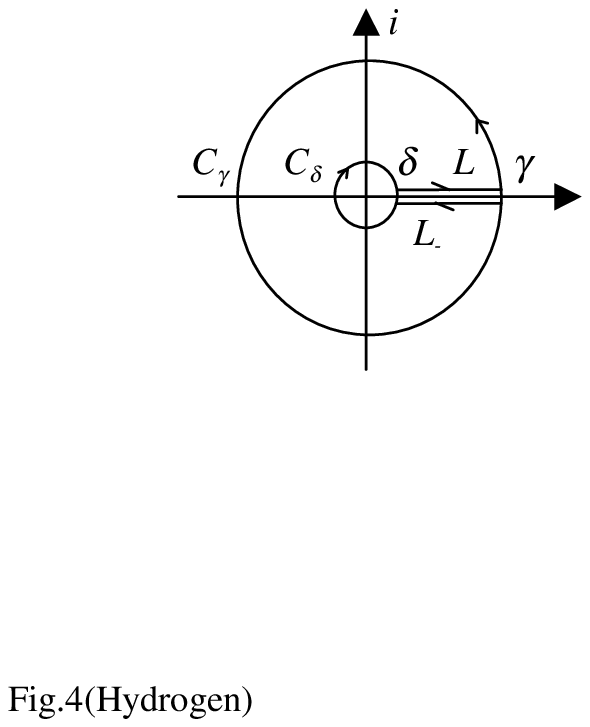}
\caption{The function has two sign-changed points $x$ and $y$.}
\label{fig5}
\end{figure}

In order to evaluate the definite integral of Eq.(\ref{s15}), consider a
contour $C$ consisting of $C_{\gamma }$, $L_{-}$, $C_{\delta }$ and $L$
around zero in the plane as shown in Fig.\ref{fig5}, the radius of circle $%
C_{\gamma }$ is large enough and the radius of circle $C_{\delta }$ is small
enough. The integrand of the following equation has no pole inside the
contour $C$, so that we have

\begin{eqnarray}
\int\nolimits_{C}\sqrt{(E+\frac{e^{2}}{z})^{2}-m_{e}^{2}c^{4}-\frac{%
j^{2}\hbar ^{2}c^{2}}{z^{2}}}dz=\nonumber \\
\int\nolimits_{C_{\gamma
}}+\int\nolimits_{L_{-}}+\int\nolimits_{C_{\delta }}+\int\nolimits_{L}=0
\label{a22}
\end{eqnarray}%
Now we evaluate the integration on each contour with our sign choice for the
double-valued function.
\begin{eqnarray}
\int\nolimits_{C_{\gamma }} &=&\int\nolimits_{C_{\gamma }}\sqrt{(E+\frac{%
e^{2}}{z})^{2}-m_{e}^{2}c^{4}-\frac{j^{2}\hbar ^{2}c^{2}}{z^{2}}}dz
\nonumber \\
&=&-\int\nolimits_{C_{\gamma }}i\sqrt{m_{e}^{2}c^{4}+\frac{j^{2}\hbar
^{2}c^{2}}{z^{2}}-(E+\frac{e^{2}}{z})^{2}}dz  \nonumber \\
&=&-i\int\nolimits_{C_{\gamma }}[\sqrt{m_{e}^{2}c^{4}-E^{2}}-\frac{Ee^{2}}{%
\sqrt{m_{e}^{2}c^{4}-E^{2}}}\frac{1}{z}+O(\frac{1}{z^{2}})]dz  \nonumber \\
&=&i\frac{2\pi iEe^{2}}{\sqrt{m_{e}^{2}c^{4}-E^{2}}}=-\frac{2\pi Ee^{2}}{%
\sqrt{m_{e}^{2}c^{4}-E^{2}}}  \label{a23}
\end{eqnarray}

\begin{eqnarray}
\int\nolimits_{C_{\delta }} &=&\int\nolimits_{C_{\delta }}i\sqrt{%
m_{e}^{2}c^{4}+\frac{j^{2}\hbar ^{2}c^{2}}{z^{2}}-(E+\frac{e^{2}}{z})^{2}}dz
\nonumber \\
&=&i\int\nolimits_{C_{\delta }}\frac{\sqrt{m_{e}^{2}c^{4}z^{2}+j^{2}\hbar
^{2}c^{2}-(Ez+e^{2})^{2}}}{z}dz  \nonumber \\
&=&i\int\nolimits_{C_{\delta }}\frac{\sqrt{j^{2}\hbar ^{2}c^{2}-e^{4}}+O(z)}{%
z}dz  \nonumber \\
&=&i(-2\pi i)\sqrt{j^{2}\hbar ^{2}c^{2}-e^{4}}\nonumber \\
&=&2\pi \sqrt{j^{2}\hbar^{2}c^{2}-e^{4}}  \label{a24}
\end{eqnarray}

\begin{figure}[htb]
\centering \includegraphics[bb=140 595 300 720,clip]{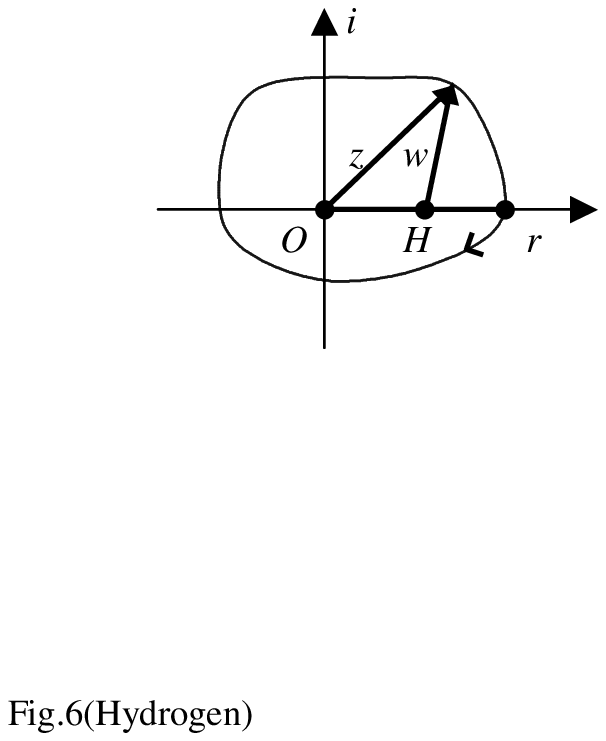}
\caption{Unit circle contour for evaluating integral}
\label{fig6}
\end{figure}

Because the integrand is a multiple-valued function, when the integral takes
over the path $L_{-}$ we have $z=e^{i2\pi }re^{0i}$, thus

\begin{equation}
\int\nolimits_{L-}=\int\nolimits_{\gamma }^{\delta }\sqrt{e^{-i2\pi }(...)}%
=-\int\nolimits_{\gamma }^{\delta }=\int\nolimits_{\delta }^{\gamma
}=\int\nolimits_{L}  \label{a25}
\end{equation}%
For a further manifestation, to define $z-H=w=\rho e^{i\beta }$, where

\begin{equation}
H=\frac{m_{e}^{2}c^{4}r^{2}+j^{2}\hbar ^{2}c^{2}-E^{2}r^{2}-e^{4}}{2Ee^{2}}
\label{a26}
\end{equation}%
we have

\begin{eqnarray}
\int\nolimits_{L-} &=&\int\nolimits_{L-}\frac{\sqrt{2Ee^{2}}\sqrt{z-H}}{z}%
dz=\int\nolimits_{L-}\frac{\sqrt{2Ee^{2}}\sqrt{w}}{z}dz  \nonumber \\
&=&\int\nolimits_{L-}\frac{\sqrt{2Ee^{2}\rho }e^{i\beta /2}}{z}dz\nonumber \\
&=&\int\nolimits_{L(\gamma \rightarrow \delta )}\frac{\sqrt{2Ee^{2}\rho }%
e^{i(\beta +2\pi )/2}}{ze^{i2\pi }}d(ze^{i2\pi })  \nonumber \\
&=&-\int\nolimits_{L(\gamma \rightarrow \delta )}\frac{\sqrt{2Ee^{2}\rho }%
e^{i\beta /2}}{z}dz\nonumber \\
&=&\int\nolimits_{L(\delta \rightarrow \gamma )}\frac{\sqrt{%
2Ee^{2}\rho }e^{i\beta /2}}{z}dz=\int\nolimits_{L}  \label{a27}
\end{eqnarray}%
where we have use the relation of $z$ and $w$ in the fourth step of the
above equation, as shown in Fig.\ref{fig6}, to note that $w$ rotates around
zero with $z$. Thus we have

\begin{eqnarray}
\int\nolimits_{L} &=&\frac{1}{2}(\int\nolimits_{L}+\int\nolimits_{L-})=-%
\frac{1}{2}(\int\nolimits_{C_{\gamma }}+\int\nolimits_{C_{\delta }})
\nonumber \\
&=&\frac{\pi Ee^{2}}{\sqrt{m_{e}^{2}c^{4}-E^{2}}}-\pi \sqrt{j^{2}\hbar
^{2}c^{2}-e^{4}}  \label{a28}
\end{eqnarray}%
Thus Eq.(\ref{s15}) becomes

\begin{eqnarray}
&&\frac{1}{\hbar c}\int\nolimits_{0}^{\infty }\sqrt{(E+\frac{e^{2}}{r}%
)^{2}-m_{e}^{2}c^{4}-\frac{j^{2}\hbar ^{2}c^{2}}{r^{2}}}dr  \nonumber \\
&=&\frac{\pi E\alpha }{\sqrt{m_{e}^{2}c^{4}-E^{2}}}-\pi \sqrt{j^{2}-\alpha
^{2}}  \label{a29}
\end{eqnarray}%
where $\alpha =e^{2}/\hbar c$ is known as the fine structure constant.

\subsection{discussion: the motion over turning points}

Following the sign change for imaginary momentum over turning points,
discussed in the preceding section, we find that the periodic condition or
standing wave condition in hydrogen atom may written as

\begin{equation}
2\int\nolimits_a^b\sqrt{\lambda -\frac{m^2}{\sin ^2\theta }}d\theta =2\pi k
\label{b71}
\end{equation}

\begin{equation}
\frac{1}{\hbar c}\int\nolimits_{x}^{y}\sqrt{(E+\frac{e^{2}}{r}%
)^{2}-m_{e}^{2}c^{4}-\frac{\lambda \hbar ^{2}c^{2}}{r^{2}}}dr=\pi s
\label{b72}
\end{equation}%
because the integrations in the regions over the turning points are
eliminated automatically. What are their physical meanings ? A direct
explanation is that it is not necessary for the electron to enter the
regions over the turning points, in compliance with classical physics.

In addition, the residue theorem we used in the paper gives out 
accurate results for our integrations, not approximate ones.


\begin{thebibliography}{9}
\bibitem{Harris} E. G. Harris, Introduction to Modern Theoretical
Physics, Vol.1\&2, John Wiley \& Sons, USA, 1975, p.263,
Eq.(10-40).

\bibitem{Harris1} See ref.\cite{Harris}, p.554, Eq.(20-9).

\bibitem{Cui} H. Y. Cui, LANL arXive.quant-ph/0102114, 2001.

\bibitem{Cui1} H. Y. Cui, in the proceedings of annual meeting of
Chinese Physical Socity, China Beijing, August, 2002.

\bibitem{Marsden} J. E. Marsden, Basic Complex Analysis, W. F.
Freeman and company, USA, 1973, p.229.

\bibitem{Schiff} L. I. Schiff, Quantum Mechanics, third ed.
McGrall-Hill, USA, 1968, p.486, Eq.(53.26).
\end{thebibliography}
\end{document}